# FRAGMENTATION MODEL OF A RAPIDLY EXPANDING RING WITH ARBITRARY CROSS-SECTION

GoloveshkinV.A. [1,2), Myagkov N.N. [1)]

[1)]*Institute of Applied Mechanics of Russian Academy of Sciences, 7 Leningradsky Prospect, Moscow 125040; e-mail: n.myagkov@iam.ras.ru, phone +7 495 946-17-65.*
[2)]*Moscow Technological University (MIREA), 20 Stromynka St., Moscow 107996 Russia.*

**Abstract**

In the paper (Goloveshkin and Myagkov 2014) we proposed a two-dimensional energy-based model of fragmentation of rapidly expanding cylinder under plane strain conditions. The model allowed one to estimate the average fragment length and the number of fragments produced by ductile fracture of the cylinder. In present note we show that the proposed approach can be used to estimate the number of fragments in a problem of fragmentation of a rapidly expanding ring with arbitrary cross-section.

*Keywords: Fragmentation model of a rapidly expanding ring; The number of fragments; Ductile fracture.*

Two one-dimensional models of the dynamic fragmentation of rapidly expanding metal cylinders have been actively discussed in recent years [1]. First is the statistics-based theory of Mott and second is the energy-based theory of Grady. The models allow one to estimate both the average fragment length and the number of fragments. It is important to note that these models are generalized to the case of the expanding rings in a trivial way due to their one-dimensionality.

In recent paper [2] we proposed a two-dimensional energy-based model of fragmentation of rapidly expanding cylinder in conditions of the ductile behavior of the material and under plane strain. The model allowed us to estimate the average fragment length and the number of fragments produced by ductile fracture of the cylinder. They obey the two-thirds power dependence on strain rate like the energy-based model of Grady. However, there is a significant difference between our model and Grady's model. The difference consists in the presence of the cylinder-wall thickness into expressions for the average fragment length and the number of fragments. In present note we show that the proposed approach [2] can be used to estimate the number of fragments in a problem of fragmentation of an expanding ring with arbitrary cross-section.

The model under consideration uses a minimum number of constants characterizing the material properties of the ring. We assume that material of the ring is incompressible with the density $\rho$ and its mechanical behavior obeys the ideal rigid-plastic model with yield stress $Y$. We also assume that the linear dimensions of the cross-section of the ring are much smaller than the ring radius $R$. The ring undergoes a uniform radial expansion with a velocity at the center of mass of the cross-section equal to $V_i$.

A key element of the problem is the model of the neck formation and evolution during rod stretching. We consider a homogeneous rod, the cross-section of which is a convex contour having a center of symmetry (it coincides with the center of mass of the cross-section). The origin of a coordinate system $(X;Y)$ lies in the geometric center of the cross-section; the axis $Z$ is parallel to the axis of the rod.

Let $\varepsilon_{xx},...\varepsilon_{yz}, \sigma_{xx},...\sigma_{yz}$ are the components of the strain rate tensor and the stress tensor, respectively. The power of internal forces per unit volume is defined as $w = \sigma_{ij}\varepsilon_{ij}$. It is easy to show that for the accepted material properties it can be written in the form

$$w = \sqrt{\frac{2}{3}} \cdot Y \sqrt{\varepsilon_{xx}^2 + \varepsilon_{yy}^2 + \varepsilon_{zz}^2 + 2\left(\varepsilon_{xy}^2 + \varepsilon_{xz}^2 + \varepsilon_{yz}^2\right)} \ . \tag{1}$$

A discontinuity of the tangential velocity equal to $[v]$ on some surface produces (according to (1)) the power of internal forces per unit surface which is equal to

$$w_\tau = \frac{1}{\sqrt{3}} Y [v] \tag{2}$$

Suppose that a continuous velocity field appeared in the rod having the form:
$u_x = u_y = 0, u_z = V_1$ at $z > \psi_1(x;y)$ and $u_x = u_y = 0, u_z = V_2$ at $z < \psi_2(x;y)$.
The velocity field between surfaces $\psi_2(x;y) < z < \psi_1(x;y)$ is arbitrary and it satisfies the continuity condition on these surfaces. We denote by $D$ a plane region in the section of the rod. Let us estimate the total power of internal forces $W$ taking into account the incompressibility condition, $\varepsilon_{xx} + \varepsilon_{yy} = -\varepsilon_{zz}$:

$$W = \iint_D dxdy \int_{\psi_2}^{\psi_1} \sqrt{\frac{2}{3}} \cdot Y \sqrt{\varepsilon_{xx}^2 + \varepsilon_{yy}^2 + \varepsilon_{zz}^2 + 2\left(\varepsilon_{xy}^2 + \varepsilon_{xz}^2 + \varepsilon_{yz}^2\right)} dz \geq \sqrt{\frac{2}{3}} \cdot Y \iint_D dxdy \int_{\psi_2}^{\psi_1} \sqrt{\varepsilon_{xx}^2 + \varepsilon_{yy}^2 + \varepsilon_{zz}^2} dz$$

$$\geq \sqrt{\frac{2}{3}} \cdot Y \iint_D dxdy \int_{\psi_2}^{\psi_1} \sqrt{\frac{3}{2}\varepsilon_{zz}^2} dz \geq Y \iint_D dxdy \int_{\psi_2}^{\psi_1} \left|\frac{\partial u_z}{\partial z}\right| dz = Y \iint_D |V_1 - V_2| dxdy = Y |V_1 - V_2| 2S \tag{3}$$

Thus, we have an estimate that gives the minimum possible power of internal forces in the problem under consideration $W_{min} = Y|V_1 - V_2|2S$, where $2S$ - is the cross-sectional area of the rod. The solution of the model problem of neck formation when the rod is stretched under the conditions of plane deformation, as described in [2], gives for the total power of the internal forces

$$W = Y|V_1 - V_2|\frac{4}{\sqrt{3}}S \qquad (4)$$

Note that the difference between (4) and the minimum possible power (3) is only ~15%.

Consider a rod in which the following velocity distribution is given at the initial instant:

$$u_x = 0, u_y = 0, \quad u_z = V_1 \text{ at } 0 < z < l_1; \quad u_x = 0, u_y = 0, \quad u_z = V_2 \text{ at } -l_2 < z < 0. \qquad (5)$$

We assume that a plane velocity field perpendicular to the X-axis analogous to the velocity field, which we considered in [2], is formed in the rod. This assumption, generally speaking, is valid if the rod cross-section is strongly elongated along the X - axis, i.e. aspect ratio $a/h \gg 1$ (see Fig. 1). However, the distinction of the solution (4) (taken from [2]) from the minimum possible power (3) by only 16% tells us that we do not make a big mistake taking the above velocity field for the case when aspect ratio $a/h \sim 1$.

We introduce a function $s(y)$ that has the meaning of the current area of the deformable part of the rod (neck), cut off by a straight line parallel to the X - axis from the cross-section of the rod (Fig. 1). It is assumed that $0 < y < h$, where $h$ is the ordinate of the furthest point of the right half of the rod section from the X- axis.

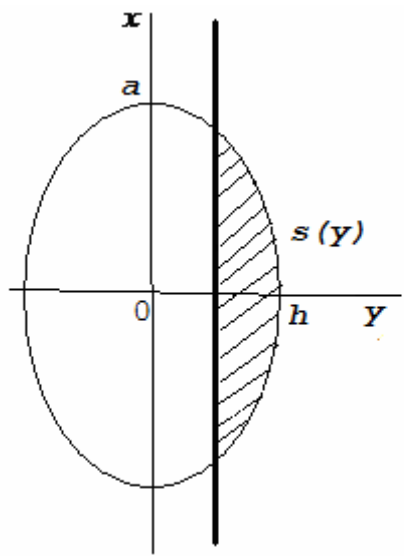

Fig. 1. The geometric meaning of the function $s(y)$.

We believe that at the current moment the deformed zone has shifted by an amount $y$. Then the value of the power of internal forces $N$ is

$$N = Y(V_1 - V_2)\frac{4}{\sqrt{3}} s(y), \tag{6}$$

where for definiteness we put $V_1 > V_2$. From the law of momentum conservation it follows that

$$l_1 \frac{dV_1}{dt} + l_2 \frac{dV_2}{dt} = 0 \tag{7}$$

From the law of energy conservation it follows that

$$\frac{dE}{dt} + N = 0 \tag{8}$$

where $E = \rho S l_1 V_1^2 + \rho S l_2 V_2^2$ is the kinetic energy of the rod. From (8) it follows that

$$2\rho S l_1 V_1 \frac{dV_1}{dt} + 2\rho S l_2 V_2 \frac{dV_2}{dt} = -Y(V_1 - V_2)\frac{4}{\sqrt{3}} s(y) \tag{9}$$

$$\frac{dy}{dt} = \frac{V_1 - V_2}{2} \tag{10}$$

Initial conditions at $t = 0$: $V_1 = V_{10}, V_2 = V_{20}, y = 0$. From (7, 9, 10) we obtain

$$4\rho S \frac{l_1 l_2}{l_1 + l_2} \frac{d^2 y}{dt^2} = -Y \frac{4}{\sqrt{3}} s(y) \tag{11}$$

with initial conditions at $t = 0$: $y = 0, \frac{dy}{dt} = \frac{V_{10} - V_{20}}{2}$. Integrating (11) we obtain

$$2\rho S \frac{l_1 l_2}{l_1 + l_2} \left(\frac{dy}{dt}\right)^2 - 2\rho S \frac{l_1 l_2}{l_1 + l_2} \left(\frac{V_{10} - V_{20}}{2}\right)^2 = -Y \frac{4}{\sqrt{3}} \int_0^y s(z) dz \tag{12}$$

A complete break of the rod occurs when $y = h$. Consequently, the condition

$$2\rho S \frac{l_1 l_2}{l_1 + l_2} \left(\frac{V_{10} - V_{20}}{2}\right)^2 \geq Y \frac{4}{\sqrt{3}} \int_0^h s(z) dz \tag{13}$$

must be satisfied for breaking the rod.

Let the moment $t = \tau$ correspond to the moment of discontinuity. Integrating power (6) with respect to time, taking into account relation (10), we obtain the work $A_f$ necessary to break the rod:

$$A_f = \frac{4}{\sqrt{3}} Y \int_0^\tau (V_1 - V_2) s(y) dt = \frac{8}{\sqrt{3}} Y \int_0^h s(y) dy. \tag{14}$$

We note that the integral $\int_0^h s(y)dy$ can be represented as $S \cdot y_c$ where $y_c$ is the coordinate of the center of mass of the right half of the rod section, and $S$ is its area. It can be seen that the value of the integral depends on the selected direction of the *X*- axis. For example, let us a square with a side $2a$. If the *X*- axis is parallel to the square side, then $h = a$ and $\int_0^h s(y)dy = a^3$, if the *X*- axis is directed diagonally, then $h = \sqrt{2} a$ and $\int_0^h s(y)dy = 2\sqrt{2}a^3/3$. The choice of the direction of the *X*- axis is determined from the condition of the minimum of the integral $\int_0^h s(y)dy$. It can be shown that the value of the integral $2\sqrt{2}a^3/3$ is minimal for the case considered above when the rod has the square cross-section with the side *2a*.

Let's consider a uniform expansion of the ring with a velocity at the center of mass of the cross-section equal to $V_i$ selecting a fragment in the form of a segment of the ring with an angle *2β*. When the ring is broken down into *n* identical fragments, $\beta = \pi/n$. We assume that the linear dimensions of the ring cross-section are small compared to its initial radius *R* and β << 1. We also assume that the fracture at both ends of the segment occurs instantaneously and in the same way. Over time after the formation, the fragment will move as a rigid body with constant velocity. Using the laws of conservation of energy and momentum is easy to estimate the reduction of kinetic energy of the fragment Δ*E* as a result of the velocity equalizing over the volume of the fragment:

$$\Delta E = \frac{2}{3} \rho V_i^2 R S \beta^3 \tag{15}$$

In paper [2] it was shown that the potential energy of the fragment can be neglected compared to its kinetic energy. Therefore, the energy balance for the problem under consideration has the form of an equality between the decrease in the kinetic energy of the fragment (15) and the minimum work of formation of the breaking surface (14): $\Delta E = (A_f)_{\min}$. As a result, we obtain the desired expression for the average number of fragments:

$$n = \pi R \left( \frac{\rho \dot{\varepsilon}_\varphi^2 S}{4\sqrt{3}\, Y \left( \int_0^h s(y)dy \right)_{min}} \right)^{1/3}, \qquad (16)$$

where $\dot{\varepsilon}_\varphi = V_i/R$. Thus, the proposed model and the obtained formula (16) allow us to estimate the average number $n$ and the average length of fragments $s = 2\pi R/n$ that are formed upon rapid expansion of a ring of arbitrary cross-section.

We estimate the ratio $J_m = \left( \int_0^h s(y)dy \right)_{min} / S$ entering in (16) for two cases of the ring cross-sections:

1) Rectangle cross-section $2a \times 2h$ where $2a$ and $2h$ are the height and thick of ring, respectively:

   a) for $a \gg h$ one gets that $J_m = h/2$;

   b) for $a = h$ one gets that $J_m = h\sqrt{2}/3$;

   c) for $a \ll h$ one gets that $J_m = a/2$.

2) Circular cross-section of radius $R$: $J_m = 4R/(3\pi)$.

It is easy to see that in all cases $J_m = k_f \cdot L$ where $L$ is a linear size of the cross-section of the ring and $k_f$ is the coefficient responsible for the cross-sectional shape. The coefficient $k_f$ depends weakly on the shape of the cross section (it changes within 15% only). We also note that the case 1 (a), as expected, yields the formula (16) for the number of fragments, which coincides with that for the cylinder [2].

**Acknowledgements.** This research was partially supported by the Russian Foundation for Basic Research (project 15-01-00565).